\documentstyle[11pt,aaspp4]{article}

\def\la{\mathrel{\hbox{\rlap{\hbox{\lower4pt\hbox{$\sim$}}}\hbox{$<$}}}}
\def\ga{\mathrel{\hbox{\rlap{\hbox{\lower4pt\hbox{$\sim$}}}\hbox{$>$}}}}
\def\etal{{\it et al.\,\,}}

\slugcomment{Submitted to {\it The Astrophysical Journal (Letters)}}

\begin{document}

\title{The Host Galaxy of GRB 971214}

\author{Daniel E. Reichart}

\affil{Department of Astronomy and Astrophysics, University of Chicago, Chicago, IL 60637} 

\begin{abstract}

GRB 971214 is the third gamma ray burst for which an optical afterglow has been detected.  I show that the spectrum of this afterglow (optical and X-ray) is inconsistent with the relativistic blast-wave model unless a significant source of extinction is introduced.  Assuming that a single absorber exists at redshift $z$, I find that its $V$-band absorption magnitude {\it at this redshift} is $A_V(z) = 1.00^{+1.55}_{-0.21}$ mag where $z = 1.89^{+0.27}_{-0.69}$.  Although this redshift is only bound from below marginally, it is strongly bound from above:  $z < 2.50$ at the 3 $\sigma$ confidence level.  
This range of values for the $V$-band absorption magnitude is too high to be consistent with absorption by either intergalactic gas or the halo of a galaxy, which implies that GRB 971214 lies either within or behind a galactic disk.  Since chance alignments of galactic disks are unlikely, this absorber is probably the host galaxy of GRB 971214.  

\end{abstract}

\keywords{gamma-rays: bursts}

\section{Introduction}

On 1997 December 12, the {\it BeppoSAX} Gamma Ray Burst Monitor discovered (Heise \etal 1997) the sixth GRB for which an X-ray afterglow has been detected (Antonelli \etal 1997) and the third GRB for which an optical afterglow has been detected.  Optical measurements of the afterglow of GRB 971214 have been reported by Halpern \etal (1997), Diercks \etal (1997), Castander \etal (1997), Rhoads (1997), and Tanvir \etal (1997) for the first three nights after the GRB.  Only $R$-, $I$-, and $J$- band measurements were possible because of the full moon.  Later measurements of the afterglow were impossible because of the proximity of the moon and the declining brightness of the afterglow. 

GRB afterglows are believed to be described by the relativistic blast-wave model (e.g., M\'esz\'aros, Rees, \& Wijers 1997).  The afterglow of GRB 970228, the first GRB for which an optical afterglow was detected, was shown to be consistent with this model (Tavani 1997, Waxman 1997, Wijers, Rees, \& M\'esz\'aros 1997, Reichart 1997a, Sahu \etal 1997, Katz \& Piran 1997).  
Reichart (1997b) showed that the afterglow of GRB 970508, the second GRB for which an optical afterglow was detected, is consistent with the simplest form of this model, i.e., an isotropic blast-wave that expands into a homogeneous medium, only if an intermediate source of extinction exists at a redshift of $z = 1.09^{+0.14}_{-0.41}$.  This is consistent with the absorption and emission features observed in spectra of this afterglow, which place GRB 970508 at a redshift of $z \ge 0.835$ (Metzger et al. 1997a, 1997b).  
The $V$-band absorption magnitude that Reichart (1997b) found for this absorber, $A_V(z) = 0.24^{+0.12}_{-0.08}$ mag, however, is too low to imply that this absorber is the GRB's host galaxy.  

In \S2, I show that the reported afterglow measurements of GRB 971214 are consistent with the general form of the relativistic blast-wave model only if a significant source of extinction is introduced.  I find the magnitude of this extinction and I place upper bounds upon the redshift of this (assumed to be) single absorber.  I draw conclusions in \S3.

\section{Data Analysis \& Model Fit}   

So far, nine measurements of the afterglow of GRB 971214 have been reported:  one 2 - 10 keV X-ray measurement, three $R$-band measurements, three $I$-band measurements, and two $J$-band measurements.  These measurements are listed in Table 1 and are plotted in Figure 1.  The photometric errors of the optical measurements range from 0.1 to 0.5 mag.  Error estimates are not available for two threshold measurements.  For the purposes of the following $\chi^2$ analysis, I have adopted uncertainties of 0.5 mag for these two measurements and an uncertainty of half of this for the X-ray measurement, which also requires error quantification.  No correction for Galactic absorption is necessary because of the GRB's high Galactic latitude.  
Although this data set contains only nine points, it has the advantage that it broadly spans the times and frequencies over which observations were possible, as limited by the fullness and the proximity of the moon (\S1).

Using the $\chi^2$ statistic, I now fit the following model to these measurements:
\begin{equation}
F_\nu = F_0 \nu^a t^b - F_{ext}(\nu;A(z),z).
\end{equation}
The first term is the extinction-free prediction of the relativistic blast-wave model (e.g., M\'esz\'aros, Rees, \& Wijers 1997).  The second second term is the correction that a redshifted source of extinction introduces (Reichart 1997b).  It is given by redshifting the interstellar absorption curve (Johnson 1965, Bless \& Savage 1972) and by specifying the magnitude of the extinction, which I parameterize as the $V$-band absorption magnitude {\it at this redshift}:  $A_V(z)$.  Extinction may be ignored in the case of the 2 - 10 keV X-ray measurement of Antonelli \etal (1997).  

The best fit is:  $\log{F_0} = -8.50^{+1.82}_{-1.21}$ cgs, $a = -0.93^{+0.04}_{-0.07}$, $b = -1.22^{+0.13}_{-0.13}$, $A_V(z) = 1.00^{+1.55}_{-0.21}$ mag, and $z = 1.89^{+0.27}_{-0.69}$ ($\chi^2_{min} = 1.53$, $\nu = 4$).  The quoted uncertainties are 1 $\sigma$ confidence intervals for one interesting parameter.  That the value of $\chi^2_{min}$ is low suggests that photometric errors may have been estimated too conservatively.  The best-fit $V$-band absorption magnitude corresponds to a hydrogen column density of $\approx 1.9$ x $10^{21}$ cm$^{-2}$.  
Consequently, a significant source of extinction is required if the relativistic blast-wave model is to apply in the case of GRB 971214.  The possibility that there is no absorber, i.e., that $A_V(z) = 0$, is ruled out at the 3.4 $\sigma$ confidence level.      
The redshift of this absorber is only constrained marginally from below:  $z > 1.20$ at the 1 $\sigma$ confidence level.  However, it is strongly constrained from above:  $z < 2.34$ at the 2 $\sigma$ confidence level and $z < 2.50$ at the 3 $\sigma$ confidence level.  This is a measure of the incompatibility of the ultraviolet absorption feature of the interstellar absorption curve with the $R$-band measurements of Diercks \etal (1997) and Castander \etal (1997).  
The best fit to Equation (1) is also plotted in Figure 1 (solid lines).  The dotted lines in this figure are the best fit for the $A_V(z) = 0$ case.

If the simplest relativistic blast-wave model is assumed, i.e., an isotropic blast-wave that expands into a homogeneous medium, the best fit is: $\log{F_0} = -7.84^{+1.12}_{-0.77}$ cgs, $b = -1.39^{+0.05}_{-0.07}$, $A_V(z) = 0.88^{+0.66}_{-0.26}$ mag, and $z = 1.89^{+0.20}_{-0.60}$ ($\chi^2_{min} = 3.32$, $\nu = 5$).   In this case, the spectral power-law index is tied to the temporal decline power-law index by $a = 2b/3$ (e.g., M\'esz\'aros, Rees, \& Wijers 1997).
The best-fit $V$-band absorption magnitude corresponds to a hydrogen column density of $\approx 1.6$ x $10^{21}$ cm$^{-2}$.  The redshift of this absorber again is only bound marginally from below:  $z > 1.29$ at the 1 $\sigma$ confidence level.  However, it is also again strongly bound from above:  $z < 2.28$ at the 2 $\sigma$ confidence level and $z < 2.51$ at the 3 $\sigma$ confidence level.  For the general relativistic blast-wave model fitted to above, $a/b = 0.77^{+0.09}_{-0.10}$, which is only inconsistent with this simplest case at only the $\approx$ 1 $\sigma$ confidence level.  

\section{Discussion and Conclusions}

The ranges of values for the $V$-band absorption magnitude above, particularly in the case of the general fit, are too high to be consistent with absorption by either intergalactic gas or the halo of a galaxy.  Consequently, if the relativistic blast-wave model is correct, GRB 971214 must lie either within or behind a galactic disk.  Since chance alignments of galactic disks are unlikely, this absorber is probably the host galaxy of GRB 971214.  

A preliminary observational result hints at this finding.  
Using the Keck II 10-m telescope after the moon had ceased to be an obstacle, Kulkarni \etal (1998) report that on 1998 January 10.58 - 10.67, they measured the $R$ magnitude of the optical transient to be 25.6 mag.  They find that this is $\sim 2$ magnitudes brighter than the extrapolated magnitude of the afterglow and consequently, that they have probably detected the host galaxy (Kulkarni \etal 1998).
However, the general fit of \S2 suggests that this discrepancy is only $\sim 0.3$ mag.  Since even small zero-point errors can significantly effect either extrapolation, further observations are required to verify this finding.

Using Equation (1) and the best-fit temporal decline power-law index for the general fit of \S2, $b = -1.22$, I have scaled each of the measurements of Table 1 to its corresponding value for December 15, just shortly after the GRB.  These points define the time-independent spectrum and are plotted in Figure 2.  The best-fit optical spectrum, as well as the extinction-free component of this spectrum (the first term of Equation (1)) are also plotted in Figure 2.  Due to the significant level of extinction implied by this fit, the ultraviolet absorption feature is easily seen.  For lower values of the redshift of the absorber, this effect is amplified:  $A_V(z = 1) = 1.66^{+0.50}_{-0.49}$.  
Had higher frequency observations been possible, this redshift might have been better constrained.

In this letter, I have again demonstrated this method by which redshifts and hydrogen column densities may be found for GRBs that are associated with host galaxies, even if spectra of sufficient quality are unattainable.  For GRB 971214, I find that an absorber of $V$-band absorption magnitude $A_V(z) = 1.00^{+1.55}_{-0.21}$ mag is required if the relativistic blast-wave model is to apply.  The case of no source of extinction is ruled out at the 3.4 $\sigma$ confidence level.  This significant level of extinction suggests that GRB 971214 is associated with a host galaxy and a preliminary observational result hints at this finding.  If this absorber is indeed the host galaxy, the redshift of GRB 971214 is $z = 1.89^{+0.27}_{-0.69}$ with an upper bound of $z < 2.50$ at the 3 $\sigma$ confidence level.  

\acknowledgments
This research has been supported by NASA grant NAG5-2868.  I thank D. Q. Lamb and F. J. Castander for their advice and their expertise, particularly on the subject of sources of extinction.  

\clearpage

\begin{deluxetable}{ccccc}
\tablecolumns{6}
\tablewidth{0pc}
\tablecaption{Observations of the Afterglow of GRB 971214}
\tablehead{
\colhead{Band} & \colhead{Date\tablenotemark{a}} & \colhead{Flux/Magnitude} & \colhead{Telescope/Instrument} & \colhead{Reference}} 
\startdata
2 - 10 keV & 15.25 & 4 x 10$^{-13}$ erg cm$^{-2}$ s$^{-1}$ & BeppoSAX/MECS & Antonelli \etal 1997 \nl
$R$ & 15.51 & $22.1 \pm 0.1$ & ARC 3.5-m & Diercks \etal 1997 \nl
$R$ & 16.52 & $23.7 \pm 0.3$ & ARC 3.5-m & Diercks \etal 1997 \nl
$R$ & 17.51 & $24.4 \pm 0.5$ & ARC 3.5-m & Castander \etal 1997 \nl
$I$ & 15.47 & $21.2 \pm 0.3$ & MDM 2.4-m & Halpern \etal 1997 \nl
$I$ & 16.47 & 22.6\tablenotemark{b} & MDM 2.4-m & Halpern \etal 1997 \nl
$I$ & 17.37-17.55 & $22.9 \pm 0.4$ & KPNO 0.9-m & Rhoads 1997 \nl
$J$ & 15.51 & $20.27 \pm 0.25$ & ARC 3.5-m & Tanvir \etal 1997 \nl
$J$ & 16.45 & 21.5\tablenotemark{c} & ARC 3.5-m & Tanvir \etal 1997 \nl
\enddata
\tablenotetext{a}{1997 December 15.25 - 17.55, UT in decimal days.}
\tablenotetext{b}{Threshold detection.}
\tablenotetext{c}{Tentative threshold detection.}
\end{deluxetable}

\clearpage

\figcaption[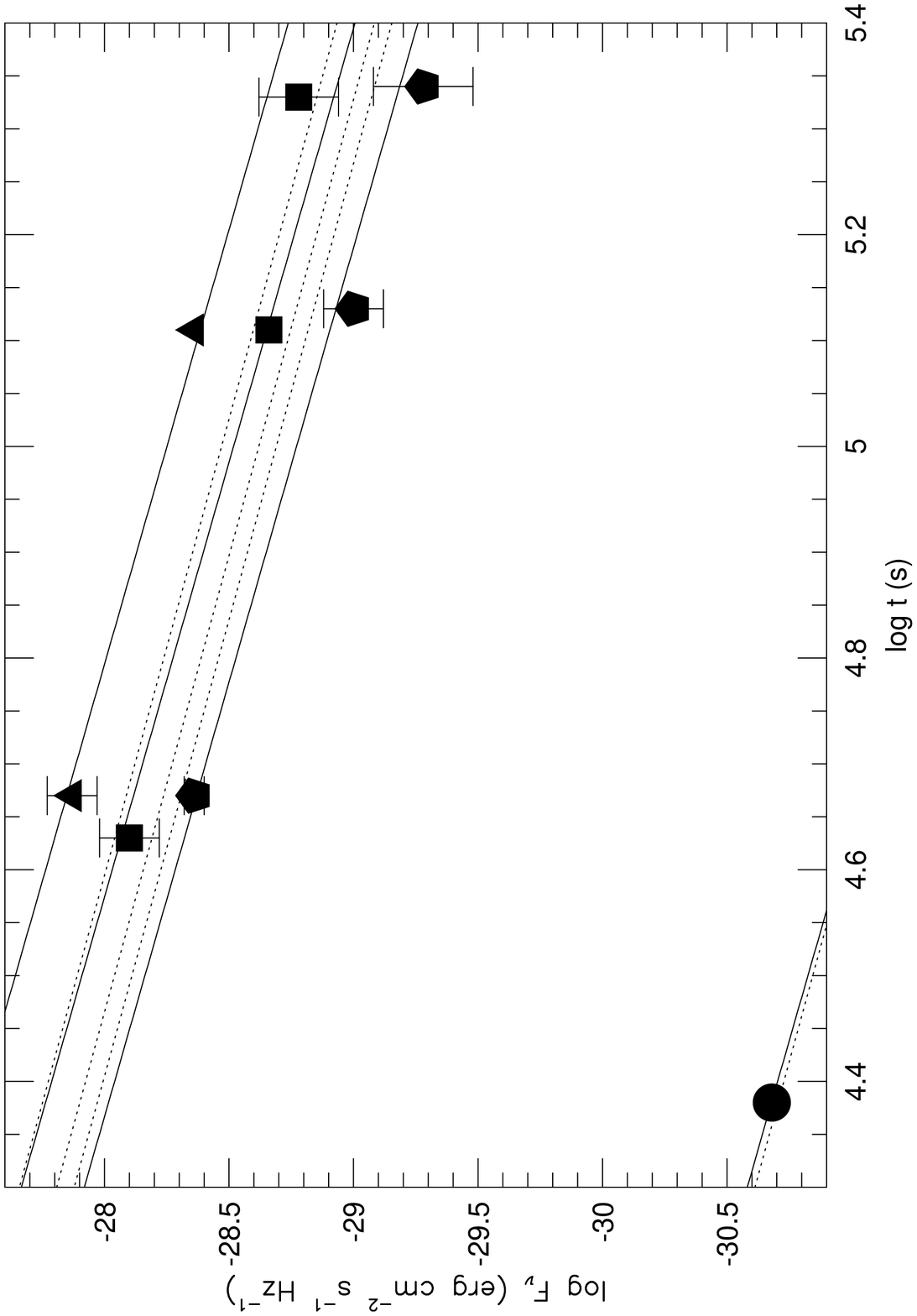]{Measurements of Table 1 and the best fit to Equation (1) (solid lines).  The circle is the 2 - 10 keV X-ray band, pentagons are the $R$ band, squares are the $I$ band, and triangles are the $J$ band.  The dotted lines are the best fit to Equation (1) if $A_V(z)$ is assumed to be negligible.\label{grbaft31.ps}}

\figcaption[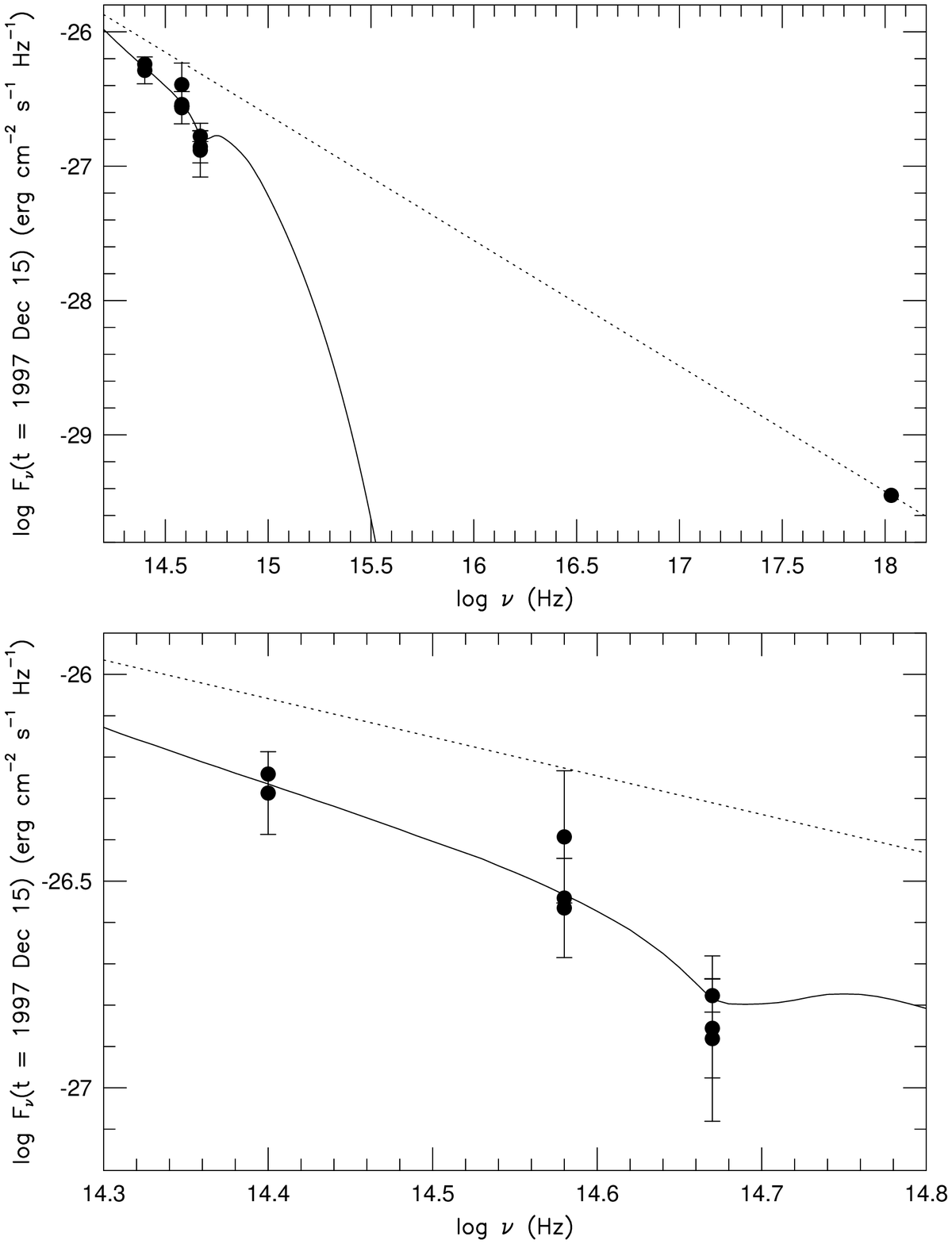]{The time-independent spectrum of the afterglow of GRB 971214 scaled to 1997 December 15.  Plotted from left to right are the $J$-, $I$-, $R$-, and 2 - 10 keV X-ray band measurements of Table 1.  The solid line is the best-fit optical spectrum and the dotted line is the extinction-free, relativistic blast-wave component of this spectrum.\label{grbaft32.ps}}  

\clearpage 

\setcounter{figure}{0}

\begin{figure}[tb]
\plotone{grbaft31.ps}
\end{figure}

\begin{figure}[tb]
\plotone{grbaft32.ps}
\end{figure}


\clearpage

\begin{thebibliography}{}
\bibitem[Antonelli \etal 1997]{aea97} Antonelli, L. A., \etal 1997, IAU Circ., 6792 
\bibitem[Bless \& Savage 1972]{bs72} Bless, R. C., \& Savage, B. D. 1972, ApJ, 171, 293
\bibitem[Castander\etal 1997]{cea97} Castander, F. J., \etal 1997, IAU Circ., 6791 
\bibitem[Diercks \etal 1997]{dea97} Diercks, A., \etal 1997, IAU Circ., 6791 
\bibitem[Halpern \etal 1997]{hea97} Halpern, J., \etal 1997, IAU Circ., 6788 
\bibitem[Heise \etal 1997]{heea97} Heise, J., \etal 1997, IAU Circ., 6787 
\bibitem[Johnson 1996]{j65} Johnson, H. L. 1965, ApJ, 141, 923
\bibitem[Katz \& Piran 1997]{kp97} Katz, J., \& Piran, T. 1997, ApJ, 490, 772
\bibitem[Kulkarni \etal 1998]{kea98} Kulkarni, S. R., \etal 1998, GCN, 27
\bibitem[M\'esz\'aros, Rees, \& Wijers 1997]{mrw97} M\'esz\'aros, P., Rees, M. J., \& Wijers, R. A. M. J., ApJ, submitted
\bibitem[Metzger \etal 1997a]{mea97a} Metzger, M. R., \etal 1997a, IAU Circ., 6676
\bibitem[Metzger \etal 1997b]{mea97b} Metzger, M. R., \etal 1997b, Nature, 387, 879
\bibitem[Reichart 1997a]{r97a} Reichart, D. E. 1997, ApJ, 485, L57
\bibitem[Reichart 1997b]{r97b} Reichart, D. E. 1997, ApJ (Letters), in press
\bibitem[Rhoads 1997]{r97} Rhoads, J. 1997, IAU Circ., 6793
\bibitem[Sahu \etal 1997]{sea97} Sahu, K., \etal 1997, Nature, 387, 476

\bibitem[Tavani 1997]{t97} Tavani, M. 1997, ApJ, 483, L87
\bibitem[Waxman 1997]{w97} Waxman, E. 1997, ApJ, 485, L5
\bibitem[Wijers, Rees, \& M\'esz\'aros 1997]{wrm97} Wijers, R. A. M. J., Rees, M. J., \& M\'esz\'aros, P. 1997, MNRAS, 288, L51

\end{thebibliography}
\end{document}